\documentclass[12pt,preprint]{aastex}
\usepackage{longtable}
\usepackage{pdflscape}	

\begin{document}

\title{A Close-in Substellar Object Orbiting the sdOB-type Eclipsing-Binary System NSVS14256825}

\author{Zhu L.-Y.\altaffilmark{1,2,3,4}, Qian S.-B.\altaffilmark{1,2,3,4}, $Fern\acute{a}ndez Laj\acute{u}s$, E.\altaffilmark{5,6}, Wang Z.-H.\altaffilmark{1,2,3,4}, Li L.-J.\altaffilmark{1,2,3}}

\singlespace

\altaffiltext{1}{Yunnan Observatories, Chinese Academy of Sciences (CAS), P. O. Box 110, 650216 Kunming, China; zhuly@ynao.ac.cn}
\altaffiltext{2}{Key Laboratory of the Structure and Evolution of Celestial Objects, Chinese Academy of Sciences, P. O. Box 110, 650216 Kunming, China}
\altaffiltext{3}{Center for Astronomical Mega-Science, Chinese Academy of Sciences, 20A Datun Road, Chaoyang District, Beijing, 100012, P. R. China}
\altaffiltext{4}{University of Chinese Academy of Sciences, Yuquan Road 19\#, Sijingshang Block, 100049 Beijing, China}
\altaffiltext{5}{Facultad de Ciencias $Astron\acute{o}micas y Geof\acute{i}sicas$, Universidad Nacional de La Plata, 1900 La Plata, Buenos Aires, Argentina}
\altaffiltext{6}{Instituto de Astrofisica de La Plata (CCT La Plata ¨C CONICET/UNLP), Paseo del Bosque s/n, La Plata, BA, B1900FWA, Argentina}

\begin{abstract}
NSVS 14256825 is the second sdOB + dM eclipsing binary systems with an orbital period of 2.65 hours. The special binary was reported to contain circumbinary planets or brown dwarfs by using the timing method. However, different results were derived by different authors because of the insufficient coverage of eclipse timings. Since 2008, we have monitored this binary for about 10 years by using several telescopes and 84 new times of light minimum in high precision were obtained. It is found that the O-C curve is increasing recently and it shows a cyclic variation with a period of 8.83 years and amplitude of 46.31 seconds. The cyclic change cannot be explained by magnetic activity cycles of the red-dwarf component because the required energy is much larger than that radiated by this component in one whole period. This cyclic change detected in NSVS1425 could be explained by the light-travel time effect via the presence of a third body. The lowest mass of the third body is determined as 14.15 $M_{jup}$ that is in the transition between planets and brown dwarfs. The substellar object is orbiting around this evolved binary at an orbital separation around 3 AU with an eccentricity of 0.12. These results indicate that NSVS 14256825 is the first sdOB-type eclipsing binary consisting of a hierarchical substellar object. The detection of a close-in substellar companion to NSVS 14256825 will provide some insights on the formation and evolution of sdOB-type binaries and their companions.

\end{abstract}


\section{Introduction}

HW Virginis (HW Vir) binaries are a group of detached eclipsing binary systems that consists a very hot subdwarf B or OB  type primary and a fully convective M-type secondary with periods usually shorter than 4 hours. They are believed to be formed through a common envelope ejection (Heber, 2009, 2016). The hot sdB-type components in this group of binaries are on the extreme horizontal branch (EHB) of the Hertzsprung-Russell diagram, burning helium in their cores and having very thin hydrogen envelopes. They have very high temperature and similar size comparing to their M-type companions, which cause the eclipse profiles of this type binaries are very sharp and deep. Benefit from this character, the light arrival time of the binary can be measured in a high precision (e.g., Kilkenny 2011, 2014). Therefore, its small wobbles caused by the orbiting of other substellar companions can be discovered. To date, this timing method is the most successful one to be applied to detect substellar objects orbiting the evolved stars, such as the giant planets orbiting around the eclipsing white dwarf binary DE CVn (Han et al., 2018) and RR Cae (Qian et al., 2012a), around eclipsing polar DP Leo (Qian et al., 2010, Beuermann et al., 2011) and HU Aqr (Qian et al. 2011, $Go\acute{z}dziewski$ et al. 2015), around eclipsing dwarf nova V2051 Oph (Qian et al., 2015) etc. Till now, the reported HW Vir-type binaries are rare with numbers not more than 20. Some of them have been found to be the hosts of substellar objects using timing method, i.e. HW Vir (Qian et al. 2008, Lee et al. 2009, Beuermann et al. 2012a), HS 0705 + 6700 (Qian et al. 2008, 2013, Beuermann et al. 2012b, ), NY Vir (Qian et al. 2012b), HS 2231+2441(Almeida et al. 2014, 2017), OGLE-GD-ECL-11388(Hong et al. 2017), 2M 1938+4603 (Baran et al. 2015) etc.

With an orbital period of 2.65 hours, NSVS 14256825 (= V1828 Aql = 2MASS J20200045 +0437564 = UCAC2 33483055 = USNO-B1.0 0946-0525128) (hereafter NSVS1425) is one of HW Vir-like eclipsing binaries possibly containing hierarchical substellar companions. Its light variability was found in the public data release from the Northern Sky Variability Survey (Wozniak et al. 2004). Wils et al. (2007) carried out the multi-band CCD observations and obtained the first group times of light minimum of NSVS1425 in a high precision (with uncertainties less than 0.0002 days). Almeida et al. (2012) analyzed their $UBVR_{c}I_{c}JH$ light curves and the radial velocity curve simultaneously using Wilson-Decinney code, and provided the reliable fundamental parameters of NSVS1425 as $M_{1}=0.419\pm0.070 M_{\odot}$, $M_{2}=0.109\pm0.023 M_{\odot}$, $R_{1}=0.188\pm0.010 R_{\odot}$, $R_{2}=0.162\pm0.008 R_{\odot}$ and $i=82^{\circ}.5\pm0^{\circ}.3$. They pointed out that NSVS1425 is the sdOB + dM eclipsing binary. Qian et al.(2010) and Zhu et al. (2011) have found the hint of the cyclic period change in this system. In 2012, Kilkenny \& Koen suggested that the period of NSVS 1425 is rapidly increasing at a rate of about $12\times10^{-12}$ days $orbit^{-1}$. Beuermann et al. (2012b) reported that there may have a giant planet with a mass of roughly 12 $M_{Jup}$ in NSVS1425. Almeida et al. (2013) revisited its O-C curve and explained the changes in O-C diagram by the presence of two circumbinary bodies with masses of 2.9 $M_{Jup}$ and 8.1 $M_{Jup}$. Wittenmyer et al. (2013) presented a dynamical analysis of the orbital stability of the model suggested by Almeida et al. (2013). They found that two-planets model in NSVS1425 is unstable on time-scales of less than a thousand years. Later, Hinse et al. (2014) concluded that the insufficient coverage of timing data prevents the reliable constrain. Recently, Nasiroglu et al.(2017) published their new times of light minimum, which extended the time span to November 2016. Their best-fitting model ruled out the two-planet model and reported a cyclic change that was explained as the presence of a brown dwarf. However, their data still do not cover a full cycle.

It is shown that the chemical compositions and evolutionary status of sdOB- and sdB-binary stars are quite different (e.g., Heber, 2016). On the T-logg diagram given by Almeida et al. (2012), the position of the primary component of NSVS1425 is close to that of the primary of the first sdOB-type binary AA Dor, but is far away from those of the other sdB-type binaries. The investigations by Kilkenny (2011, 2014) showed that the O-C curve of AA Dor is constant indicating that no substellar objects orbiting around it. Therefore, are there any substellar objects orbiting NSVS1425 is a very interesting question. In this paper, we present our 84 newly determined high precision timings for NSVS1425 obtained from the observation between Dec. 2008 to Dec. 2018, which effectively extend the baseline of the timing data and covered more than a full cycle of the cyclic variation in the O-C diagram. Combined with the high precision timings collected from the literature, we performed a new orbital period investigation of this HW-Vir type binary.

\section{Observations and Data Reduction}
We have been monitoring a group of HW Vir-like eclipsing binaries since 2006. For NSVS1425, we began to observe it in December 2008 with several small telescope in China and Argentina, i.e., the 2.4-m and 70-cm telescopes in Lijiang station, 1.0-m and 60-cm telescopes in Kunming station of Yunnan Observatories, Chinese Academy of Sciences (YNO 2.4m, YNO 70cm, YNO 1m and YNO 60cm), the 2.16-m and 85-cm telescope at Xinglong Station of National Astronomical Observatories, Chinese Academy of Sciences (NAOC 2.16m and NAOC 85cm), and the 2.15-m Jorge Sahade telescope at Complejo Astronomico El Leoncito (CASLEO 2.15m), San Juan, Argentina. These seven telescopes are all equipped with CCD cameras and the standard Johnson$-$Cousin$-$Bessel $BVR_{c}I_{c}$ filters. Using these telescopes, we obtained the observations covering ten years from December 2008 to December 2018. All image reductions were done by using the IRAF package. 72 primary eclipsing profiles and 12 secondary eclipsing profiles were obtained. They all show symmetric light variations. To derived the times of light minimum, we modeled these eclipsing profiles using the Amplitude version of Gaussian peak function:

\begin{equation}
G(t)=m_{0}+Ae^{-\frac{(t-t_{c})^{2}}{2w}}.
\end{equation}
\noindent In this function, $m_{0}$ is the offset; A is the amplitude; $t_{c}$ is the timing and the $w$ is the width with $2w=FWHM/\sqrt{ln(4)}$. Figure 1 shows some examples of the observed eclipsing light curves(open circles) and the corresponding fits(solid lines).

\begin{figure}[h!]
\begin{center}
\includegraphics[width=12cm]{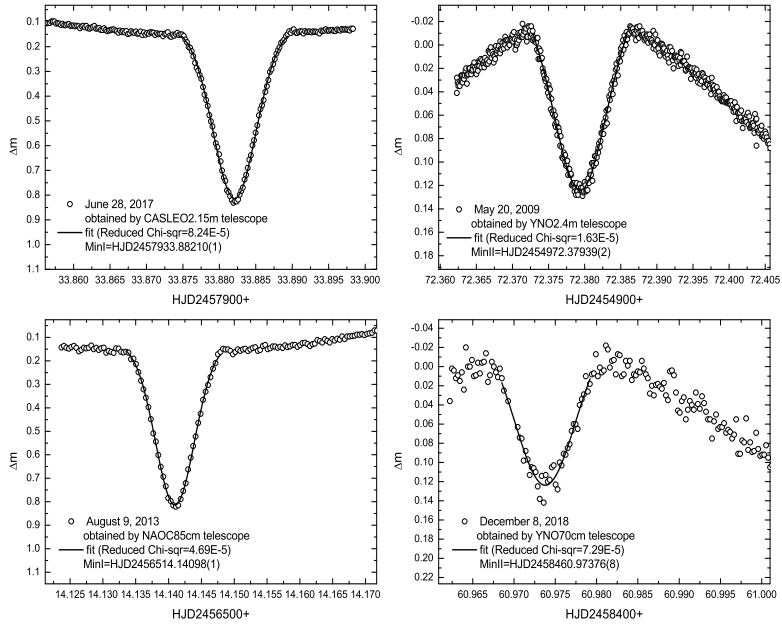}
\caption{Some eclipsing profiles of NSVS1425. Open circles are the observations and solid lines are the fit curves. The light curves shown in the upper panel were obtained by 2m class telescopes, while those in the lower panel were obtained by 1m class telescopes. Left panel shows the primary eclipsing profiles and right panel shows the secondary ones.}
\end{center}
\end{figure}

In total, 84 newly times of light minimum were determined based on our observations. We converted these eclipse times from HJD to BJD with the numerical procedure developed by Eastman et al. (2010) \footnote{http://astroutils.astronomy.ohio-state.edu/time/}, which are listed in the first and sixth columns in Table 1. The second and seventh column of Table 1 provides their errors. One can see that the uncertainties of all newly determined timings are smaller than 0.00015 days. These high precision eclipse times are very important to detect the existence of the substellar third body by using the light-travel time effect (LTT).

\begin{table}
\caption{New times of light minimum of NSVS1425.}\label{oc}
\begin{center}
\tiny
\begin{tabular}{llllllllll}
\hline
BJD                &  Err.(d)      &   E           &   O-C (d)       &  Tele.         &  BJD                &  Err.(d)      &   E           &   O-C (d)       &  Tel.     \\\hline
2454818.90501      &  0.00005      &  4935         &  -0.000258      &  NAOC $85\rm{cm}$   &  2456047.36959      &  0.00001      &  16065        &  0.000235       &  YNO $60\rm{cm}$   \\
2454933.36268      &  0.00005      &  5972         &  -0.000562      &  YNO $60\rm{cm}$   &  2456069.33405      &  0.00004      &  16264        &  0.000243       &  YNO $60\rm{cm}$   \\
2454936.34294      &  0.00008      &  5999         &  -0.000403      &  YNO $60\rm{cm}$   &  2456069.38926      &  0.00008      &  16264.5      &  0.000266       &  YNO $60\rm{cm}$   \\
2454961.39796      &  0.00003      &  6226         &  -0.000311      &  YNO $1.0\rm{m}$   &  2456164.14541      &  0.00003      &  17123        &  0.000218       &  NAOC $85\rm{cm}$   \\
2454963.38463      &  0.00007      &  6244         &  -0.000376      &  YNO $60\rm{cm}$   &  2456219.11183      &  0.00002      &  17621        &  0.000322       &  YNO $1.0\rm{m}$     \\
2454968.24111      &  0.00003      &  6288         &  -0.000358      &  YNO $60\rm{cm}$   &  2456234.01231      &  0.00003      &  17756        &  0.000294       &  YNO $60\rm{cm}$   \\
2454969.23445      &  0.00008      &  6297         &  -0.000385      &  YNO $60\rm{cm}$   &  2456248.96791      &  0.00008      &  17891.5      &  0.000199       &  YNO $60\rm{cm}$   \\
2454969.28957      &  0.00013      &  6297.5       &  -0.000452      &  YNO $60\rm{cm}$   &  2456249.02323      &  0.00004      &  17892        &  0.000332       &  YNO $60\rm{cm}$   \\
2454972.38013      &  0.00002      &  6325.5       &  -0.000367      &  YNO $2.4\rm{m}$   &  2456373.41486      &  0.00013      &  19019        &  0.000308       &  YNO $60\rm{cm}$   \\
2454986.23200      &  0.00004      &  6451         &  -0.000446      &  YNO $2.4\rm{m}$   &  2456440.30158      &  0.00007      &  19625        &  0.000306       &  YNO $1.0\rm{m}$     \\
2454994.28935      &  0.00006      &  6524         &  -0.000413      &  YNO $60\rm{cm}$   &  2456456.30576      &  0.00003      &  19770        &  0.000236       &  YNO $1.0\rm{m}$     \\
2455000.24974      &  0.00005      &  6578         &  -0.000226      &  YNO $60\rm{cm}$   &  2456514.14177      &  0.00001      &  20294        &  0.000223       &  NAOC $85\rm{cm}$   \\
2455031.26470      &  0.00010      &  6859         &  -0.000397      &  YNO $1.0\rm{m}$   &  2456533.12611      &  0.00003      &  20466        &  0.000192       &  YNO $1.0\rm{m}$     \\
2455104.11174      &  0.00004      &  7519         &  -0.000284      &  YNO $2.4\rm{m}$   &  2456557.07731      &  0.00003      &  20683        &  0.000206       &  YNO $1.0\rm{m}$     \\
2455118.01888      &  0.00001      &  7645         &  -0.000289      &  YNO $2.4\rm{m}$   &  2456758.39958      &  0.00001      &  22507        &  0.000060       &  YNO $2.4\rm{m}$   \\
2455118.07402      &  0.00003      &  7645.5       &  -0.000336      &  YNO $2.4\rm{m}$   &  2456823.18914      &  0.00003      &  23094        &  0.000004       &  NAOC $85\rm{cm}$   \\
2455118.12921      &  0.00002      &  7646         &  -0.000328      &  YNO $2.4\rm{m}$   &  2456908.17717      &  0.00006      &  23864        &  -0.000048      &  YNO $1.0\rm{m}$     \\
2455146.05390      &  0.00001      &  7899         &  -0.000294      &  YNO $2.4\rm{m}$   &  2456960.05296      &  0.00001      &  24334        &  -0.000100      &  YNO $2.4\rm{m}$   \\
2455153.00750      &  0.00004      &  7962         &  -0.000264      &  NAOC $85\rm{cm}$   &  2456962.03971      &  0.00009      &  24352        &  -0.000084      &  YNO $2.4\rm{m}$   \\
2455165.03824      &  0.00002      &  8071         &  -0.000305      &  YNO $2.4\rm{m}$   &  2456966.01314      &  0.00003      &  24388        &  -0.000123      &  YNO $1.0\rm{m}$     \\
2455296.38336      &  0.00011      &  9261         &  -0.000400      &  YNO $60\rm{cm}$   &  2456972.96670      &  0.00001      &  24451        &  -0.000133      &  NAOC $85\rm{cm}$   \\
2455333.35878      &  0.00004      &  9596         &  -0.000314      &  YNO $2.4\rm{m}$   &  2456973.96007      &  0.00001      &  24460        &  -0.000131      &  NAOC $85\rm{cm}$   \\
2455334.35217      &  0.00001      &  9605         &  -0.000291      &  YNO $2.4\rm{m}$   &  2457187.25783      &  0.00008      &  26392.5      &  -0.000384      &  NAOC $85\rm{cm}$   \\
2455380.15742      &  0.00002      &  10020        &  -0.000305      &  YNO $1.0\rm{m}$   &  2457278.04048      &  0.00003      &  27215        &  -0.000458      &  NAOC $85\rm{cm}$   \\
2455437.11053      &  0.00013      &  10536        &  -0.000256      &  YNO $60\rm{cm}$   &  2457933.82780      &  0.00004      &  33156.5      &  -0.001046      &  CASLEO $2.15\rm{m}$  \\
2455438.10395      &  0.00007      &  10545        &  -0.000203      &  YNO $60\rm{cm}$   &  2457933.88289      &  0.00001      &  33157        &  -0.001143      &  CASLEO $2.15\rm{m}$  \\
2455444.17447      &  0.00002      &  10600        &  -0.000260      &  NAOC $2.16\rm{m}$  &  2457991.60849      &  0.00001      &  33680        &  -0.001216      &  CASLEO $2.15\rm{m}$  \\
2455445.05746      &  0.00002      &  10608        &  -0.000263      &  YNO $1.0\rm{m}$   &  2457991.66373      &  0.00003      &  33680.5      &  -0.001163      &  CASLEO $2.15\rm{m}$  \\
2455450.02436      &  0.00005      &  10653        &  -0.000199      &  NAOC $85\rm{cm}$   &  2458008.16467      &  0.00002      &  33830        &  -0.001156      &  YNO $2.4\rm{m}$   \\
2455453.11482      &  0.00002      &  10681        &  -0.000215      &  NAOC $85\rm{cm}$   &  2458011.14475      &  0.00002      &  33857        &  -0.001178      &  YNO $2.4\rm{m}$   \\
2455499.03051      &  0.00002      &  11097        &  -0.000163      &  NAOC $85\rm{cm}$   &  2458019.53317      &  0.00001      &  33933        &  -0.001183      &  CASLEO $2.15\rm{m}$  \\
2455688.32227      &  0.00004      &  12812        &  -0.000034      &  NAOC $85\rm{cm}$   &  2458019.58832      &  0.00004      &  33933.5      &  -0.001230      &  CASLEO $2.15\rm{m}$  \\
2455692.29572      &  0.00003      &  12848        &  -0.000053      &  NAOC $85\rm{cm}$   &  2458019.64351      &  0.00001      &  33934        &  -0.001227      &  CASLEO $2.15\rm{m}$  \\
2455700.35324      &  0.00012      &  12921        &  0.000156       &  YNO $60\rm{cm}$   &  2458287.79745      &  0.00005      &  36363.5      &  -0.001233      &  CASLEO $2.15\rm{m}$  \\
2455721.32418      &  0.00011      &  13111        &  0.000011       &  YNO $60\rm{cm}$   &  2458287.85272      &  0.00002      &  36364        &  -0.001150      &  CASLEO $2.15\rm{m}$  \\
2455737.32837      &  0.00001      &  13256        &  -0.000047      &  YNO $2.4\rm{m}$   &  2458288.73570      &  0.00002      &  36372        &  -0.001163      &  CASLEO $2.15\rm{m}$  \\
2455783.35435      &  0.00004      &  13673        &  -0.000089      &  YNO $2.4\rm{m}$   &  2458366.05278      &  0.00006      &  37072.5      &  -0.001164      &  YNO $60\rm{cm}$   \\
2455784.12698      &  0.00008      &  13680        &  -0.000078      &  NAOC $85\rm{cm}$   &  2458366.10801      &  0.00004      &  37073        &  -0.001121      &  YNO $60\rm{cm}$   \\
2455798.14458      &  0.00002      &  13807        &  0.000008       &  YNO $2.4\rm{m}$   &  2458451.97914      &  0.00002      &  37851        &  -0.001070      &  YNO $70\rm{cm}$   \\
2455872.09538      &  0.00002      &  14477        &  0.000141       &  YNO $2.4\rm{m}$   &  2458460.97453      &  0.00008      &  37932.5      &  -0.001173      &  YNO $70\rm{cm}$   \\
2455879.04895      &  0.00001      &  14540        &  0.000141       &  YNO $2.4\rm{m}$   &  2458461.02974      &  0.00005      &  37933        &  -0.001150      &  YNO $70\rm{cm}$   \\
2455883.02224      &  0.00006      &  14576        &  -0.000038      &  YNO $60\rm{cm}$   &  2458462.02315      &  0.00002      &  37942        &  -0.001108      &  YNO $70\rm{cm}$   \\
\hline
\end{tabular}
\end{center}
\end{table}

\section{Orbital Period Investigation}
Based on the times of light minimum, several authors have studied the O-C diagram of NSVS1425. Among them, Nasiroglu et al.(2017) collected the most comprehensive timings till the end of 2016. By adding our new eclipse times, we calculated the cycle number E and the O-C values according to the following ephemeris from Beuermann et al. (2012b).

\begin{equation}
Min.I (BJD) = 2454274.208923(4)+0^{d}.1103741324(3)\times{E}\label{ep1},
\end{equation}

The new constructed O-C curve extends the coverage baseline for another two years, which is plotted in Fig. 2. Our O-C values calculated from new timings are shown with green dots, while others are displayed with blue dots. From this Figure, one can see that the another minimum in the O-C curve has been caught, making the analysis based on one full cycle of the periodical variation possible. As the LTT signals caused by the low mass or/and long period objects are small. Low precision timings may lead to this signal submerged in the scattering points, such as the case of EG Cep (Zhu et al., 2009). Therefore, we just use the timings with errors smaller than 0.0002 to reconstructed the O-C curve and display it in the upper panel of Fig 3. The cyclic variation is obvious in this Figure. To describe the O-C curve, the following equation was used:
\begin{eqnarray}
O-C=\Delta T_{0} + \Delta P_{0}\times E + \tau,
\end{eqnarray}\noindent where $\Delta T_{0}$ and $\Delta P_{0}$ are, respectively
the revised epoch and period with respect to the ephemeris values in
Eq \ref{ep1}. $\tau$ is described by Irwin (1952) as the cyclic change term due to the light-time effect caused by a third body:

\begin{eqnarray}
\tau &=& K[(1-e^{2})\frac{sin(\nu+\omega)}{1+ecos\nu}+esin\omega]\nonumber \\
     &=& K[\sqrt{1-e^{2}}sinE^{*}cos\omega+cosE^{*}sin\omega]\label{eq2}.
\end{eqnarray}
\noindent In this equation, e is the eccentricity, $\nu$ is the true
anomaly, $\omega$ is the logitude of the periastron passage in the
plane of the orbit, and $E^{*}$ is the eccentric anomaly.
$K=a\textrm{sin}i^{'}/c$ is the projected semi-major axis given in days,
where $a$ is the semi-major axis of elliptic orbit. By means of the Levenberg-Marquardt
method, the results of the nonlinear fit for the O-C diagram are
obtained. The solid line in the top panel of Fig. 3 represents a combination of a revised linear ephemeris and a periodic variation. All derived parameters are listed in Table 2. The revised period is $0.1103741030(5)$ days.

Our final result indicates that the $O-C$ curve shows a cyclic change with a period of $8.83(6)$\,years and an amplitude of $0.000536(5)$ \,days that could be seen more clearly in the middle panel of Fig. 3. The residuals are shown in the bottom panel and its standard deviation is 0.00005. The results show that the third body in NSVS1425 has a minimal mass of 14.15 $M_{Jup}$, which orbits the central HW Vir type binary in an eccentric orbit with $e\sim0.12(2)$.

\begin{figure}[h!]
\begin{center}
\includegraphics[width=12cm]{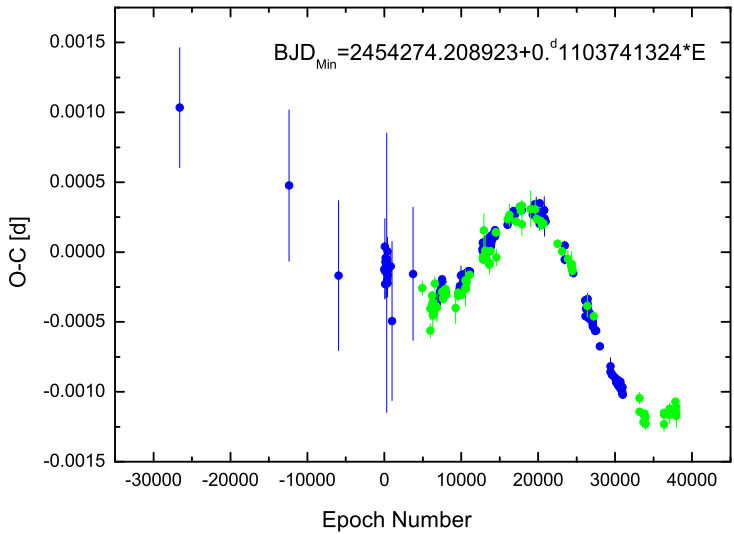}
\caption{The O-C diagrams of the HW Vir-type binary NSVS1425 constructed based on all available timings. Green dots are derived from our new data. }
\end{center}
\end{figure}

\begin{figure}[h!]
\begin{center}
\includegraphics[width=12cm]{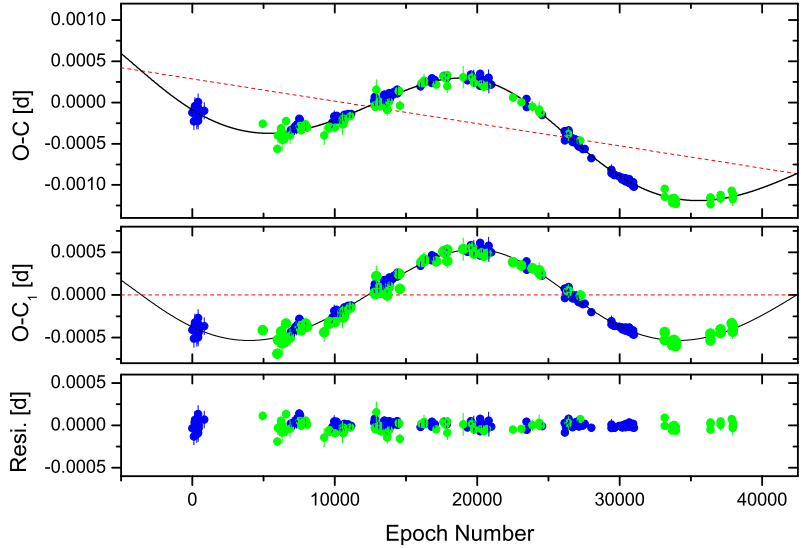}
\caption{The O-C diagrams of the HW Vir-type binary NSVS1425 constructed by the high precision timings. Green dots refer to the newly data obtained by us, while blue dots to the others collected from the literature. The solid line in the top panel represents a combination of a revised linear ephemeris (the dashed red line) and a periodic variation (also shown as the solid line in the middle panel where the $(O-C)_{1}$ values were calculated with respect to the revised linear ephemeris). The residuals are displayed in the lowest panel where no variations can be traced there.}
\end{center}
\end{figure}

\begin{table}[h!]
\caption{Orbital parameters of NSVS1425 and its circumbinary substellar object.}\label{para}
\begin{center}
\scriptsize
\begin{tabular}{ll}
\hline
Parameters                                            &Values:  \\\hline
Correction on the initial epoch, ${T_{0}}$ (days)           & $ 2.87(\pm0.10)\times10^{-4}$          \\
Revised period, $P$ (days)                                  & $0.1103741030(\pm0.0000000005)$        \\
Light travel-time effect amplitude, $K$ (days)                & $0.000536(\pm0.000005)$  \\
Eccentricity, e                                              & $0.12(\pm0.02)$  \\
Orbital period, $P_3$ (years)                                  & $8.83(\pm0.06)$                      \\
Longitude of the periastron passage, $\omega$ (deg)            & $133.3(\pm10.3)$   \\
Periastron passage, T (HJD)                                    & $2456816.0(\pm93.7)$   \\
Projected semi-major axis, $a_{12}\sin{i^{'}}$ $(AU)$            & $0.0928(\pm0.0009)$  \\
Mass function,$f(m)$ $(M_{\odot})$                                  & $9.63(\pm0.32)\times10^{-6}$         \\
Mass of the third body, $M_{3}$ $(M_{Jup}, {i'=90^{\circ}}$)           & $14.15(\pm0.16)$                        \\
Orbital separation, $d_{3}$ ($AU,{i'=90^{\circ}}$)              & $3.12(\pm0.07)$                        \\
\hline
\end{tabular}
\end{center}
\end{table}

\section{Discussions and conclusions}

NSVS 1425 is the second HW Vir type eclipsing binaries consisting of a very hot subdwarf OB type primary and a fully convective M-type secondary. Its sdOB primary lies in the EHB of the Hertzsprung-Russell diagram, having a very high temperature about 40000K (Almeida et al. 2013) and small size with radius of 0.19 $R_{\odot}$, while its M dwarf secondary has very low temperature and comparative size with sdOB primary. thanks to these characters, the eclipsing profiles produced by them are very sharp and deep, which makes the light arrival time of the central object can be measured in a high precision and therefore its small wobbles caused by the orbiting of the substellar companions can be discovered. As the high surface gravities and the compact structures of sdB, sdOB or white dwarfs, both the radial velocity and transit methods (have been extensively used to search for planets around solar-type main-sequence stars) have a low efficiency to detect substellar companions (exoplants or brown dwarfs) to those post-red giant branch stars. Instead, this timing method base on the light-travel time effect (LTT) is the most successful one to be applied to find substellar objects orbiting these evolved stars. It is similar to the radio approach used to discovery planets around pulsars(Wolszczan \& Frail, 1992; Backer et al., 1993).

We have monitored NSVS 1425 for ten years and obtained 84 high precision new times of light minimum, which effectively extended the baseline of the data and covered another minimum in the O-C curve, making a full cycle constructed by the high precision data available. Based on these data, we reanalyzed the O-C diagram and found a cyclic oscillation with an amplitude of 0.00053 days (or 45.8 seconds) and a period of 8.83 years, which can be explained by the wobble of the binary's barycenter via the existence of a third body. Our updated parameters confirmed that there is a hierarchical substellar object orbiting around the central sdOB + dM type eclipsing binary with orbital eccentricity of 0.12. The relationship between the mass ($M_{3}$) and distance at periastron ($d_{3}$) varying along with the inclination of tertiary component ($i^{'}$) are shown in Fig 4. The distance at periastron ($d_{3}$) of this tertiary component ranges from 3.26 AU to 3.12 AU as its inclination $(i^{'})$ varies from 10 degree to 90 degree. That means this tertiary component is a close-in object. when the orbital plane of this third body is coplanar with the orbiting plane of the inner binary, its mass ($M_{3}$) should be $14.3 M_{Jup}$, which lies in the boundary zone between planets and brown dwarfs.

\begin{figure}[!h]
\begin{center}
\includegraphics[width=12cm]{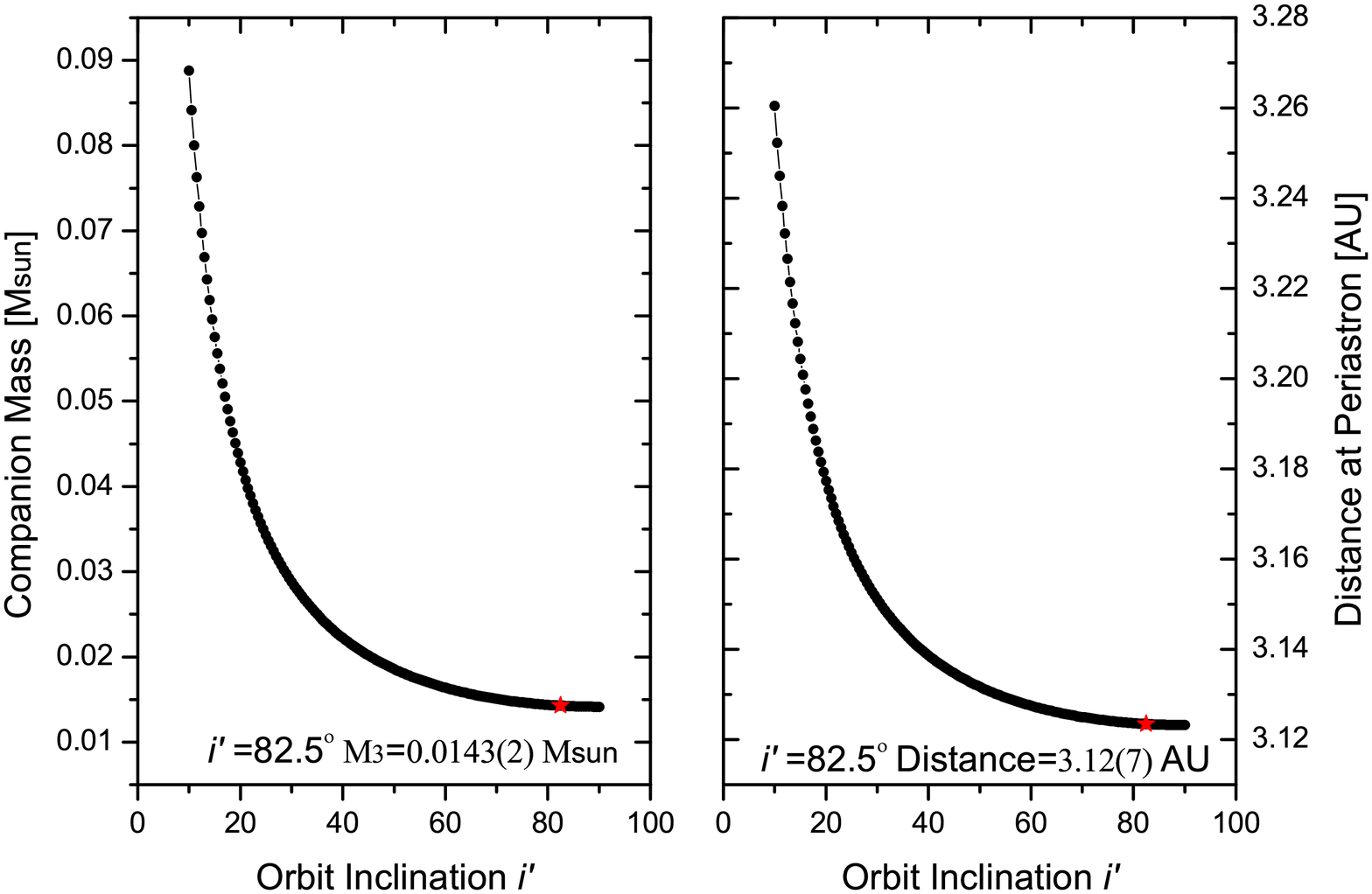}
\caption{Relations between the mass $M_{3}$ and the distance at periastron $d_{3}$ of the third body and its orbital inclination $i^{'}$ in the NSVS1425 system. The red stars represent the value When its orbital plane is coplanar with the orbiting plane of the inner binary.}
\end{center}
\end{figure}

As suggested by many authors, the cyclic period oscillations of close binaries comprised of cool star components also can be explained by magnetic activity cycles (Applegate 1992). For NSVS1425, the secondary component is a fully convective red-dwarf star with mass of 0.109 $M_{\odot}$ and effective temperature of 2550 K (Almeida et al. 2012). In order to check this possibility, we computed the energies required to cause this cyclic period change using the same method proposed by Brinkworth et al. (2006) and plotted its relationship with the assumed shell mass of the secondary in Fig. 4 (solid line). The dashed line in Fig. 4 represents the total energy that radiates from the fully convective red-dwarf secondary in one whole period (8.83 years). One can see that the required energies are much larger than the total energy radiated from secondary star in one whole period. Therefore, the cyclic change can not be explained by magnetic activity cycles of the secondary component in NSVS1425.

\begin{figure}[h!]
\begin{center}
\includegraphics[width=12cm]{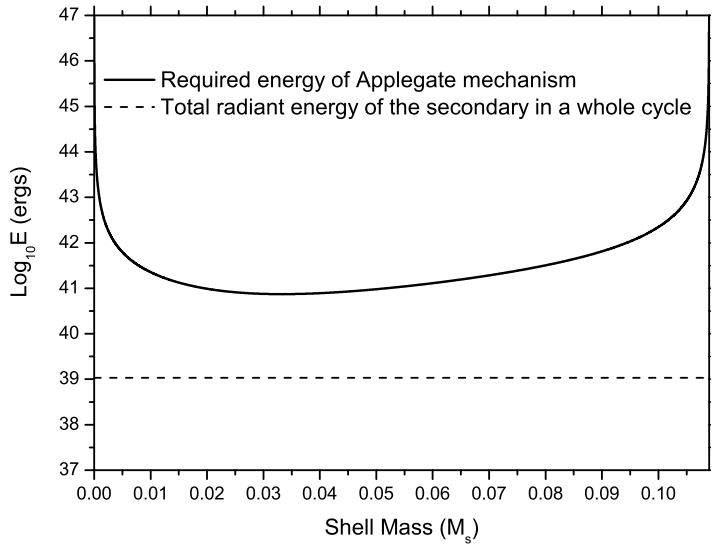}
\caption{Energy required to produce the cyclic oscillation in the O-C diagrams using Applegate's mechanism (solid line). $M_{s}$ is the assumed shell
mass of the cool component. The dashed line represents the total energy that radiates from the fully convective red-dwarf secondary in one whole period $\thicksim$ 8.83 years.}
\end{center}
\end{figure}

Some sdB eclipsing binaries have been reported with substellar objects orbiting around them, but sdOB eclipsing binaries not. NSVS1425 is the first one. For single sdB- and sdO-stars, their chemical compositions and evolutionary status are quite different. sdB stars are mostly helium poor, while helium abundances of sdO stars show a variety of helium abundances (e.g., Heber 2016). Moreover, most sdB stars were found in binary systems, while the binary frequency of sdO stars is very low. These properties may indicate that their formations are different. As for the two sdOB eclipsing binary stars, AA Dor and NSVS1425, although they are located in the same region on the T-logg diagram (e.g., Almeida et al. 2012), their properties on the tertiary companions are quite different. A close-in substellar object is orbiting to NSVS1425, while no substellar objects are orbiting around AA Dor (e.g., Kilkenny 2011, 2014). If the circumbinary substellar object in NSVS1425 is formed from the remaining common-envelope material (second generation planets), why there are no substellar objects orbiting around AA Dor? Maybe, AA Dor and NSVS1425 have different formation channels. The detection of the close-in substellar object orbiting around NSVS1425 will provide us more information on the formation and evolution of sdOB-type binaries and their companions.

\acknowledgments{This work is partly supported by the National Natural Science Foundation of China (Nos. 11573063), the Key Science Foundation of Yunnan Province (No. 2017FA001), CAS "Light of West China" Program and CAS Interdisciplinary Innovation Team. New CCD photometric observations of NSVS14256825 were obtained with the 2.16m and 85-cm telescope at Xinglong observational station of NAOC, the 2.4m, 1.0m, 60cm and 70cm telescopes at YNOs, and the 2.15m telescope at Complejo Astronomico El Leoncito (CASLEO), San Juan, Argentina.}

\end{document}